\documentclass[conference]{IEEEtran}
\IEEEoverridecommandlockouts

\usepackage{cite}
\usepackage{amsmath,amssymb,amsfonts}
\usepackage{algorithm}
\usepackage{algorithmic}
\usepackage{graphicx}
\usepackage{textcomp}
\usepackage{xcolor}
\usepackage{etoolbox}
\def\BibTeX{{\rm B\kern-.05em{\sc i\kern-.025em b}\kern-.08em
    T\kern-.1667em\lower.7ex\hbox{E}\kern-.125emX}}
\usepackage{fancyhdr}

\begin{document}
\fancypagestyle{firstpageheader}{
  \fancyhf{} 
  \renewcommand{\headrulewidth}{0pt}
  \fancyhead[C]{\footnotesize Results in part accepted for presentation at the 2026 IEEE Consumer Communications and Networking Conference (CCNC),\\ to be held in Las Vegas, USA, Jan. 2026}
}
\pagestyle{plain}

\title{Optimizing Multi-UAV 3D Deployment for Energy-Efficient Sensing over Uneven Terrains}

\vspace{-0.8cm}

\author{
    \IEEEauthorblockN{
        Rushi Moliya\IEEEauthorrefmark{1}, 
        Dhaval K. Patel\IEEEauthorrefmark{1}, 
        Brijesh Soni\IEEEauthorrefmark{2}, 
        Miguel L\'opez-Benítez\IEEEauthorrefmark{3}\IEEEauthorrefmark{4}
    } 
    
    \IEEEauthorblockA{\IEEEauthorrefmark{1}School of Engineering and Applied Science, Ahmedabad University, Ahmedabad 380009, India} 
    \IEEEauthorblockA{\IEEEauthorrefmark{2}Department of Computer Science and Engineering, The Ohio State University, Columbus, OH 43210, USA} 
    \IEEEauthorblockA{\IEEEauthorrefmark{3}School of Computer Science and Infomatics, University of Liverpool, Liverpool L69 3DR, United Kingdom} 
    \IEEEauthorblockA{\IEEEauthorrefmark{4}ARIES Research Centre, Antonio de Nebrija University, 28015 Madrid, Spain} 
    
    \IEEEauthorblockA{
        Emails:  \{rushi.m, dhaval.patel\}@ahduni.edu.in, soni.152@osu.edu, m.lopez-benitez@liverpool.ac.uk
    }
}

\vspace{-0.8cm}

\maketitle

\thispagestyle{firstpageheader}

\begin{abstract}
In this work, we consider a multi-unmanned aerial vehicle (UAV) cooperative sensing system where UAVs are deployed to sense multiple targets in terrain-aware line of sight (LoS) conditions in uneven terrain equipped with directional antennas. To mitigate terrain-induced LoS blockages that degrade detection performance, we incorporate a binary LoS indicator and propose a bounding volume hierarchy (BHV)-based adaptive scheme for efficient LoS evaluation. We formulate a bi-objective problem that maximizes the probability of cooperative detection with minimal hover energy constraints governing spatial, orientational, and safety constraints. To address the problem, which is inherently non-convex, we propose a hierarchical heuristic framework that combines exploration through a genetic algorithm (GA) with per-UAV refinement via particle swarm optimization (PSO), where a penalty-based fitness evaluation guides solutions toward feasibility, bounded within constraints. The proposed methodology is an effective trade-off method of traversing through a complex search space and maintaining terrain-aware LoS connectivity and energy aware deployment. Monte Carlo simulations on real-world terrain data show that the proposed GA+PSO framework improves detection probability by 37.02\% and 36.5\% for 2 and 3 UAVs, respectively, while reducing average excess hover energy by 45.0\% and 48.9\% compared to the PSO-only baseline. Relative to the non-optimized scheme, it further achieves 59.5\% and 54.2\% higher detection probability with 59.8\% and 65.9\% lower excess hover energy, thereby showing its effectiveness with a small number of UAVs over uneven terrain.

\end{abstract}

\begin{IEEEkeywords}
Unmanned aerial vehicles, Cooperative spectrum sensing, Eigenvalue-based detection, Uneven terrain, Genetic algorithm, Particle swarm optimization.
\end{IEEEkeywords}
\vspace{-0.25cm}
\section{Introduction}
\vspace{-0.05cm}

Multi-UAV systems have been widely implemented in civilian, commercial, and especially military contexts in recent years \cite{Li2025, Wei2024}. Their agility, flexible deployment, and ability to maintain reliable LoS links make UAVs highly suitable for wireless communication and sensing tasks, especially spectrum sensing. Due to their scalability and mobility, multi-UAV networks increase detection in large and inaccessible areas, and cooperative spectrum sensing (CSS) provides a strong alternative to stationary terrestrial schemes \cite{Zeng2016}.

Energy detection (ED) has been used extensively in spectrum sensing because it is simple and has been extended to crowdsourced sensing \cite{Bhattacharya2020} and 3D RF sensor networks \cite{Kleber2022}. Nonetheless, the sensitivity of ED to noise uncertainty leads to the well-known signal-to-noise ratio (SNR) wall, causing severe degradation in dynamic low-SNR environments. To address this, a robust alternative, eigenvalue-based detection (EBD) has been suggested in the \cite{Moliya2025}, which does not depend on prior knowledge of noise variance, but is resilient against uncertainty.  

Despite these advances, important limitations remain. Many deployment and trajectory models assume fixed UAV altitudes and simplify the problem to two dimensions (2D), which is impractical since real UAVs operate in 3D environments where altitude critically impacts coverage \cite{Moliya2025}, sensing accuracy, and energy consumption. Furthermore, several studies implicitly assume uninterrupted LoS, neglecting terrain obstructions such as hills or buildings \cite{Moliya2025, Kanzariya2025}. Such assumptions are unrealistic in urban or hilly environments and lead to infeasible deployment strategies in practice. Without explicit consideration of terrain-aware LoS, deployment strategies risk being infeasible in real-world scenarios \cite{Savkin2025}.

We overcome these limitations by proposing a novel LoS-aware 3D UAV deployment model in CSS over uneven terrain. In the proposed system, the terrain-induced LoS is explicitly accounted as a binary indicator, and the realistic energy costs are also considered. The joint objective is formulated as maximizing the sum of detection probability and minimizing average excess hover energy, hence ensuing both sensing reliability and sustainable UAV operation. To achieve efficient LoS verification, we propose a scalable adaptive sampling approach using BVH that reduces computational cost while maintaining accuracy. To tackle the resulting non-convex and discontinuous optimization problem, we develop a hierarchical heuristic framework that combines GA exploration with per-UAV PSO refinement. A penalty-based fitness function also imparts spatial, orientation and safety constraints, thereby making its deployment reliable as well as practical. The main contributions of this paper are summarized as follows.

\begin{figure*}[t]
    \centering
    \includegraphics[width=1\textwidth]{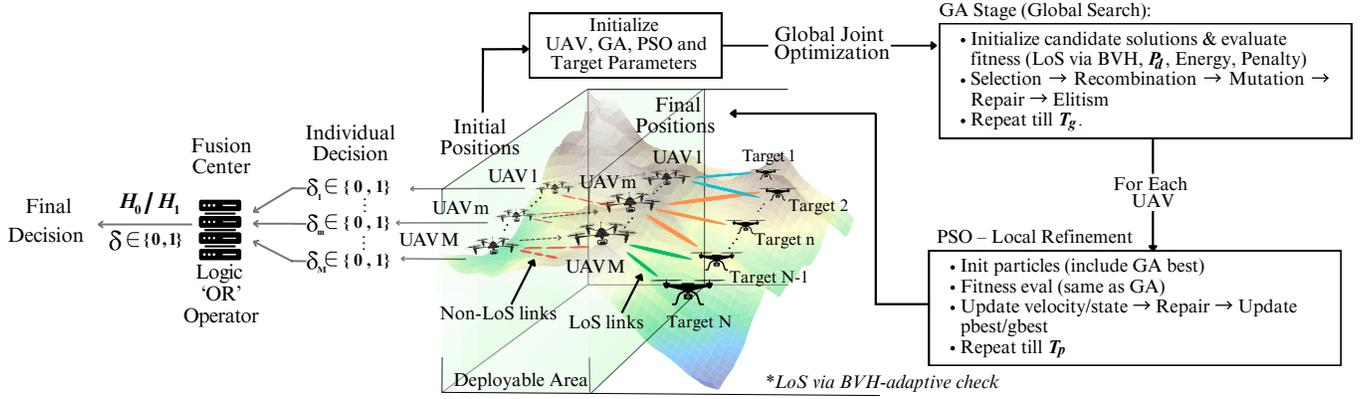}
    \vspace{-0.8cm}
    \caption{Multi-UAV cooperative sensing network over real terrain, where $M$ UAVs detect $N$ targets using eigenvalue-based detection along with the hierarchical GA+PSO optimization flow.}
    \label{fig:UAV_scenario}
    \vspace{-0.4cm}
\end{figure*}

\begin{itemize}
    \item We present a novel LoS-aware 3D UAV deployment framework for multi-UAV CSS over uneven terrain, jointly optimizing placement and antenna orientation under a bi-objective that maximizes detection probability and minimizes hover energy.   
    
    \item We propose a scalable BVH-based adaptive method for efficient binary LoS determination and develop a hierarchical heuristic framework that integrates global GA exploration with local PSO refinement, guided by a penalty-based fitness function to handle complex spatial, orientational, and safety constraints.  
    
    \item We validate the framework through Monte Carlo simulations on real-world 3D terrain data, demonstrating improved detection performance and reduced energy consumption over existing baselines.  
\end{itemize}

\vspace{-0.1cm}

\section{System Model}
As illustrated in Fig.~\ref{fig:UAV_scenario}, we consider a centralized multi-UAV CSS network operating in a three-dimensional environment with uneven terrain and terrain-induced LoS constraints. The system deploys $M$ UAVs, indexed by $m \in \{1,\dots,M\}$, to cooperatively detect $N$ targets, indexed by $n \in \{1,\dots,N\}$, while minimizing energy expenditure. The setup adopts a 3D Cartesian coordinate system, where the UAV position set is defined as ${\mathcal{P}}=\{\mathbf{p}_1,\dots,\mathbf{p}_M\}$ with $\mathbf{p}_m=[x_m,y_m,z_m]^{\mathsf T}$ denoting the coordinates of UAV $m$, and the target position set as $\mathcal{P}^t=\{\mathbf{p}_1^t,\dots,\mathbf{p}_N^t\}$ with $\mathbf{p}_n^t=[x_n^t,y_n^t,z_n^t]^{\mathsf T}$ denoting the coordinates of target $n$. The UAVs deployment is carried out in a terrain-aware environment that explicitly accounts for elevation variations and obstruction effects. The terrain elevation is modeled as a weighted sum of $G$ two-dimensional Gaussian components,
\vspace{-0.2cm}
\begin{align}
\beta_{\text{terrain}}(x,y)=\sum_{i=1}^{G} h_i \exp\!\left(-\Big[\tfrac{(x-\mu_{x,i})^2}{2\sigma_{x,i}^2}+\tfrac{(y-\mu_{y,i})^2}{2\sigma_{y,i}^2}\Big]\right), 
\vspace{-0.15cm}
\end{align}
\noindent where $h_i$ is the peak height, $(\mu_{x,i},\mu_{y,i})$ the center, and $(\sigma_{x,i},\sigma_{y,i})$ the spatial spreads of the $i$th Gaussian. To capture terrain masking, we define the binary-LoS indicator between two 3D coordinates  $\mathbf{p}_1$ and $\mathbf{p}_2$ as
\vspace{-0.1cm}
\begin{align}
\small
\mathcal{I}(\mathbf{p}_1,\mathbf{p}_2)=
\begin{cases}
1, & z(t)>\beta_{\text{terrain}}(x(t),y(t)),~\forall t\in[0,1],\\
0, & \text{otherwise},
\end{cases}
\end{align}
where $\mathbf{p}(t)=(1-t)\mathbf{p}_1+t\mathbf{p}_2$ parameterizes the line segment, and $z(t)$ denotes the altitude component along this path. For brevity, let $\mathcal{I}_{mn}=\mathcal{I}(\mathbf{p}_m,\mathbf{p}_n^t)$ denote the LoS condition between UAV $m$ and target $n$. Each UAV carries a steerable directional antenna, and both positions and boresight orientations are centrally coordinated to maintain alignment under LoS constraints, forming the basis for the sensing and energy models that follow.
\vspace{-0.1cm}
\subsection{Spectrum Sensing Model}
Each UAV conducts spectrum sensing to detect transmissions from assigned targets. Detection performance depends jointly on LoS availability and antenna alignment, while cooperative processing across UAVs enhances reliability through spatial diversity.  

\subsubsection{Eigenvalue-Based Detection (EBD)}
According to \cite{Zeng2007Eigenvalue}, EBD exploits the distribution of eigenvalues in the sample covariance matrix to detect signals based on statistical variations, remaining robust under noise uncertainty since it does not rely on prior knowledge of noise variance. The detection problem is formulated as a binary hypothesis test, where $H_0$ denotes signal absence and $H_1$ its presence:
\vspace{-0.1cm}
\begin{align}
\begin{cases}
H_0: & \mathbf{y}(k) = \mathbf{n}(k), \\
H_1: & \mathbf{y}(k) = \mathbf{x}(k) + \mathbf{n}(k),
\end{cases}
\quad k=0,1,\dots,K-1,
\vspace{-0.1cm}
\end{align}

\noindent where $\mathbf{y}(k)\in\mathbb{C}^{L\times 1}$ denotes the received vector at a UAV with $L$ sensing elements, $K$ is the number of samples, $\mathbf{x}(k)$ is the signal component with covariance $E[\mathbf{x}(k)\mathbf{x}^H(k)]=\sigma_s^2\mathbf{I}$, and $\mathbf{n}(k)$ is AWGN with zero mean and covariance $E[\mathbf{n}(k)\mathbf{n}^H(k)]=\sigma_n^2\mathbf{I}$.

The sample covariance is given by $\mathbf{R}_y=\tfrac{1}{K}\sum_{k=0}^{K-1}\mathbf{y}(k)\mathbf{y}^H(k)$, with eigenvalues $\lambda_1\geq\lambda_2\geq\dots\geq\lambda_L$. The test statistic is $T=\lambda_1/\lambda_L$, compared against threshold $\rho_{\text{EBD}}$. For $K\gg L$, $T$ is approximately Gaussian: $T\sim\mathcal{N}(1,2/K)$ under $H_0$ and $T\sim\mathcal{N}(1+\text{SNR},\,2(1+\text{SNR})^2/K)$ under $H_1$, where $\text{SNR}=\sigma_s^2/\sigma_n^2$. From, this false-alarm and detection probabilities are given by~\cite{Zeng2007Eigenvalue},
\vspace{-0.1cm}
\begin{align}
\footnotesize
P_{fa} = Q\!\left(\frac{\rho_{\text{EBD}}-1}{\sqrt{2/K}}\right),
P_d = Q\!\left(\frac{\rho_{\text{EBD}}-(1+\text{SNR})}{\sqrt{2(1+\text{SNR})^2/K}}\right),
\vspace{-0.1cm}
\end{align}
\noindent where $Q(x)=\tfrac{1}{\sqrt{2\pi}}\int_x^{\infty}e^{-t^2/2}\,dt$. For a specified $P_{fa}$, the threshold is
\begin{equation}
\rho_{\text{EBD}}=1+\sqrt{\tfrac{2}{K}}\,Q^{-1}(P_{fa}),
\end{equation}
and substituting this in (4) yields the closed-form detection probability
\vspace{-0.1cm}
\begin{equation}
P_d=Q\!\left(\frac{Q^{-1}(P_{fa})-\text{SNR}\cdot\sqrt{K/2}}{1+\text{SNR}}\right).
\end{equation}
\vspace{-0.1cm}

\subsubsection{Directional Antenna Model}
In multi-UAV cooperative sensing, each UAV employs a directional antenna to concentrate energy toward intended targets, thereby improving received signal strength and mitigating off-boresight interference. The main lobe is characterized by azimuth and elevation half-power beamwidths $(\alpha_a,\alpha_e)$, with negligible gain outside this region~\cite{Miron2006}. Terrain-induced visibility is incorporated via the LoS indicator $\mathcal{I}_{mn}=I(\mathbf{p}_m,\mathbf{p}_n^t)\in\{0,1\}$ defined in (2), so that link gain is counted only under LoS conditions. Following~\cite{Miron2006}, the directional gain between UAV $m$ and target $n$ is \vspace{-0.2cm}
\begin{equation}
G^{lin}_{m,n} = 
\begin{cases} \mathcal{I}_{mn}\cdot \Psi^{lin}_{m,n} (\eta_m, \zeta_m), & |\eta_{m,n} - \eta_m| \leq \alpha_a \ \text{and} \\& |\zeta_{m,n} - \zeta_m| \leq \alpha_e, \\
0, & \text{otherwise}.
\end{cases}
\footnotesize 
\vspace{-0.1cm}
\end{equation} 
\noindent where $\Psi^{lin}_{m,n} = \exp \left( -\frac{ (\eta_{m,n} - \eta_m)^2 }{2 \sigma_\eta^2} - \frac{ (\zeta_{m,n} - \zeta_m)^2 }{2 \sigma_\zeta^2 } \right)$, the azimuth angle is $\eta_{m,n} = \arctan \left( \frac{y_n^t - y_m}{x_n^t - x_m} \right)$ and elevation angle is $\zeta_{m,n} = \arctan \left( \frac{z_n^t - z_m}{\sqrt{(x_n^t - x_m)^2 + (y_n^t - y_m)^2}} \right)$ from the UAV $m$ to the target $n$. The angular spreads $\sigma_\eta$ and $\sigma_\zeta$ are related to the half-power beamwidths by $\sigma_\eta=\alpha_a/(2\sqrt{2\ln2})$ and $\sigma_\zeta=\alpha_e/(2\sqrt{2\ln2})$. The resulting signal-to-interference-plus-noise ratio (SINR) for UAV $m$ sensing target $n$ with transmit power $P_n$ is given as in \cite{Xue2018},
\vspace{-0.1cm}
\begin{equation}
\gamma_{m,n} = \frac{P_n \cdot \beta_0 \cdot N_t \cdot G^{lin}_{m,n}}{d_{m,n}^2 \cdot \sigma_n^2},
\vspace{-0.1cm}
\end{equation}
\noindent where $d_{m,n}$ is the Euclidean distance, $\beta_0$ is the reference channel gain, $N_t$ is the number of antenna elements, and $\sigma_n^2$ is the noise variance. Substituting $\gamma_{m,n}$ into (6) gives the per-link detection probability,
\vspace{-0.1cm}
\begin{equation}
P_d = Q\!\left(\frac{Q^{-1}(P_{fa})-\gamma_{m,n}\sqrt{K/2}}{1+\gamma_{m,n}}\right).
\end{equation}
\vspace{-0.1cm}

\subsubsection{Cooperative Sensing Model}
Each UAV is equipped with a sensing front-end capable of monitoring only a limited frequency band, denoted by $(f_{\min,m},f_{\max,m})$. Targets transmit on discrete channels selected from their allocated sets $\mathcal{F}_n=\{f_{n,1},f_{n,2},\dots,f_{n,|\mathcal{F}_n|}\}$, where $|\mathcal{F}_n|$ is the number of channels available to target $n$. For a given frequency $f\in\mathcal{F}_n$, the individual detection probability of UAV $m$ for target $n$ is
\vspace{-0.1cm}
\begin{equation}
\footnotesize
P^f_{m,n}=
\begin{cases}
Q\!\left(\frac{Q^{-1}(P_{fa})-\gamma_{m,n}\sqrt{K/2}}{1+\gamma_{m,n}}\right), & f\in(f_{\min,m},f_{\max,m}),\\
0, & \text{otherwise}.
\end{cases}
\end{equation}
\vspace{-0.3cm}

To enhance reliability,  local decisions $\delta_m\in(0,1)$ from multiple UAVs are fused at a centralized controller, where $\delta_m=1$ indicates a detection at the $m^{th}$ UAV. An OR-rule is adopted as a simple scheme that maximizes detection probability, especially under blocked or low-SNR conditions. Thereby, cooperative detection probability for target $n$ on channel $f$ is given as
\vspace{-0.1cm}
\begin{equation}
P^f_n(\mathbf{p},\eta,\zeta)=1-\prod_{m=1}^M\!\left(1-P^f_{m,n}(\mathbf{p}_m,\eta_m,\zeta_m)\right).
\end{equation}
\vspace{-0.1cm}

The overall cooperative sensing performance of the multi-UAV system is defined as its sum detection probability, 
\vspace{-0.35cm}

\begin{equation}
P_{\text{sum}}=\sum_{n=1}^N\sum_{f \in \mathcal{F_\text{$n$}}} P^f_n(\mathcal{P},\eta,\zeta).
\end{equation}

\vspace{-0.25cm}
\subsection{Energy Consumption Model}
In multi-UAV systems with limited battery capacity, energy-aware deployment is critical for sustaining long-duration missions. For static hovering, propulsion power dominates consumption and increases with altitude due to reduced air density and higher induced power requirements. The relative altitude of the $m^{th}$ UAV is $h_m \triangleq z_m - \beta_{\text{terrain}}(x_m,y_m)$, and its hover energy is modeled as~\cite{Leishman2006},
\begin{equation}
\small
E_{\text{hover},m} = P_0 \cdot t_d \cdot \left(1-\frac{h_m}{H_s}\right)^{-2.128}, \quad 0 < h_m < H_s,
\end{equation}
where $P_0$ denotes the nominal hover power at ground level, $H_s$ is the atmospheric scale height, and $t_d$ is the hover duration. To capture sensitivity in the low-altitude regime, we define the average excess hover energy relative to the safety floor altitude $H_{\text{safe}}$, which represents the minimum operational height maintained above the terrain to ensure safe UAV flight, as
\vspace{-0.4cm}
\begin{equation}
E_{\text{avg,ex}} = \frac{1}{M}\sum_{m=1}^M \big( E_{\text{hover}}(h_m) - E_{\text{hover}}(H_{\text{safe}}) \big).
\vspace{-0.2cm}
\end{equation}



\subsection{Problem Formulation}
\vspace{-0.1cm}

The multi-UAV deployment problem seeks to maximize sum detection probability while minimizing hovering energy consumption over uneven terrain. The joint objective is optimized over UAV positions $\mathcal{P}$ and antenna orientations $(\boldsymbol{\eta},\boldsymbol{\zeta})$ subject to safety and operational constraints, as
\vspace{-0.2cm}
\begin{align}
\vspace{-0.2cm}
& \max_{\mathbf{P}, \boldsymbol{\eta}, \boldsymbol{\zeta}} \quad \lambda_S \cdot P_{\text{sum}} - \lambda_E \cdot E_{\text{avg,ex}} \tag{15} \\
& \text{s.t.} \nonumber \\
& \mathbf{p}_m \in \mathcal{D}, \quad \forall m \in \{1, \dots, M\}, \tag{16a} \\
& \| \mathbf{p}_m - \mathbf{p}_n^t \|_2 \geq S_{\min}, \quad \forall m,n, \tag{16b} \\
& \| \mathbf{p}_m - \mathbf{p}_l \|_2 \geq R_{\min}, \quad \forall m \neq l, \tag{16c} \\
& -\pi \leq \eta_m \leq \pi, \quad -\tfrac{\pi}{2} \leq \zeta_m \leq \tfrac{\pi}{2}, \quad \forall m, \tag{16d} \\
& \beta_{\text{terrain}}(x_m,y_m) + H_{\text{safe}} \leq z_m \leq H_{\max}, \quad \forall m. \tag{16e}
\vspace{-0.2cm}
\end{align}
In this formulation, (16a) confines UAVs to the deployable region, (16b) enforces a minimum separation from targets, (16c) ensures safe spacing among UAVs, (16d) limits antenna orientation, and (16e) guarantees terrain-aware altitude safety. The weights $\lambda_S$ and $\lambda_E$ trade off sensing accuracy and energy usage, adjustable per objective needs.

\vspace{-0.1cm}

\section{Proposed Methodology}
\vspace{-0.1cm}

\subsection{ Proposed BVH-Based Adaptive LoS Determination}
\vspace{-0.1cm}

Efficient LoS determination is critical for cooperative sensing since detection performance depends on the binary indicator $\mathcal{I}(\mathbf{p}_m,\mathbf{p}_n^t)$. A naive approach would sample $T$ uniformly spaced points along each UAV–target path, requiring $\mathcal{O}(TG)$ operations for $G$ Gaussian terrain components. To achieve scalability, the proposed method integrates a BVH with adaptive sampling.
\vspace{-0.4cm}

\begin{algorithm}[H]
\caption{BVH-Based Adaptive LoS Determination}
\label{algo:LoSAlgo}
\textbf{Initialize:} UAV position $\mathbf{p}_1$, target position $\mathbf{p}_2$, BVH root \\ \hspace{3em} $\mathcal{R}$, terrain function $\beta_{\mathrm{terrain}}$, tolerance $\varepsilon$\\
1: \textbf{if} projection does not intersect $\mathcal{R}$  \textbf{then} \\
2:\hspace{2em} \textbf{return} $\mathcal{I}(\mathbf{p}_1,\mathbf{p}_2)\gets 1$ \hfill \% Early acceptance\\
3: Initialize $\mathcal{T}\gets\{0,1\}$\\
4: \textbf{while} $\exists\,(t_i,t_j)\in\mathcal{T};\ |t_i-t_j|>\varepsilon$ \textbf{do}\\
5:\hspace{1em} $t_{m}\gets (t_i+t_j)/2$, $\mathbf{p}_{mid}\gets (1-t_{m})\mathbf{p}_1+t_{m}\mathbf{p}_2$\\
6:\hspace{1em} \textbf{if} $(x_{mid},y_{mid})$ in BVH \textbf{then}\\
7:\hspace{2em} $z_{\text{terr}}\gets\beta_{\text{terrain}}(x_{mid},y_{mid})$\\
8:\hspace{2em} \textbf{if} $z_{mid}\leq z_{\text{terr}}$ \textbf{then return} $\mathcal{I}(\mathbf{p}_1,\mathbf{p}_2)\gets 0$ \hfill \% Blocked\\
9:\hspace{0.8em}\textbf{end if}; Add $t_{mid}$ to $\mathcal{T}$\\
10:\textbf{end while}\\
11:\textbf{return} $\mathcal{I}(\mathbf{p}_1,\mathbf{p}_2)\gets 1$ \hfill \%LoS presence
\end{algorithm}
\vspace{-0.4cm}

Each Gaussian component is enclosed in an axis-aligned bounding box, $\mathcal{B}_i=[\mu_{x,i}-r_{x,i},\mu_{x,i}+r_{x,i}] \times [\mu_{y,i}-r_{y,i},\mu_{y,i}+r_{y,i}]$, where $r_{x,i}=k_o\sigma_{x,i}$ and $r_{y,i}=k_o\sigma_{y,i}$ are coverage radii determined by scaling factor \(k_o\). Recursive partitioning along the longest axis yields a balanced BVH that excludes non-intersecting terrain in $\mathcal{O}(\log G)$ time.

Algorithm~\ref{algo:LoSAlgo} summarizes the proposed BVH-based Adaptive LoS Determination algorithm. If the UAV–target path does not intersect the BVH root, LoS is accepted (Line-2). Otherwise, sampling along the path $\mathbf{p}(t)=(1-t)\mathbf{p}_1+t\mathbf{p}_2$, $t\in[0,1]$ starts with $\{0,1\}$ and iteratively bisects intervals (Line-5). When a midpoint lies in a BVH node, terrain elevation is checked: if $z(t_m)\leq \beta_{\text{terrain}}(x(t_m),y(t_m))$, LoS is blocked (Line-8); otherwise, it is retained (Line-9). The process continues until intervals are below $\varepsilon$, after which LoS is confirmed (Line-11).

\begin{figure*}[t]
    \centering
    \includegraphics[width=0.9\textwidth]{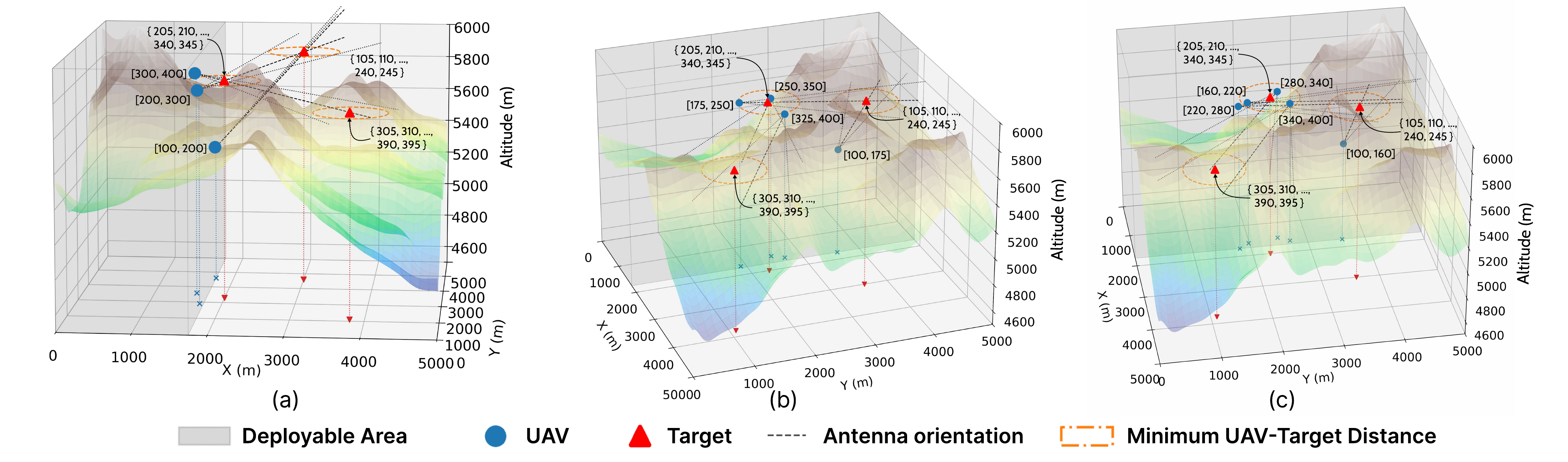}
    \vspace{-0.3cm}
    \caption{3D Deployment results in multiple UAVs perceive multiple target scenarios over real SRTM-based uneven terrain: (a) $M = 3$, (b) $M = 4$, (c) $M = 5$}
    \label{fig:UAV_Deployment_GA+PSO}
    \vspace{-0.4cm}
\end{figure*}

The BVH pruning ensures that only potentially obstructing terrain regions are evaluated, while adaptive refinement focuses sampling near critical points. This hybrid approach achieves $\mathcal{O}(T' + \log G)$ complexity, where $T'$ is the number of midpoint evaluations and $T'\ll T$ in practice, resulting in a substantial improvement over uniform sampling.

\vspace{-0.1cm}

\subsection{Proposed Optimization Framework}

\vspace{-0.1cm}

The optimization problem in (15)–(16e) is highly non-convex due to spatial, angular, and terrain-aware altitude constraints, while the binary LoS indicator introduces discontinuities that make gradient-based solvers ineffective and prone to local minima. To address this, we employ a two-stage hierarchical metaheuristic: a global GA to explore the deployment space, followed by per-UAV PSO refinement to improve individual placements under fixed neighbors. The procedure is summarized in Algorithm~\ref{algo:ga_pso} and detailed below. 

\textbf{GA Stage:}
Here, $N_g$ candidate solutions are initialized, where each solution encodes a complete multi-UAV deployment vector with $\mathcal{P}$ and $(\boldsymbol{\eta},\boldsymbol{\zeta})$ (Line–2). For solution $i$ at iteration $t$, the deployment is represented as
$ \mathbf{g}^{(t)}_{i} = \big[\,x^{(t)}_{i,1},y^{(t)}_{i,1},z^{(t)}_{i,1},\eta^{(t)}_{i,1},\dots,y^{(t)}_{i,M},z^{(t)}_{i,M},\eta^{(t)}_{i,M},\zeta^{(t)}_{i,M}\,\big],
\label{eq:ga-gene} $
where $(x^{(t)}_{i,m},y^{(t)}_{i,m},z^{(t)}_{i,m})$ denote the 3D coordinates of UAV $m$, and $(\eta^{(t)}_{i,m},\zeta^{(t)}_{i,m})$ are its azimuth and elevation angles, initialized within the specified constraints. The fitness of a solution $\mathbf{g}^{(t)}_{i}$ is evaluated using a scalarized objective function with explicit penalties for constraint violations.

\begin{algorithm}[H]
\caption{Hierarchical GA+PSO Optimization Framework}
\textbf{Initialize:} GA: $ N_g$, $ T_g$, $ p_{\text{mut}}$, $ \Delta_j$, $ \mu_{\text{elite}}$, $ l$; PSO: $ w_{\max}$, $ w_{\min}$, $ c_1$, $ c_2$, $ N_p$, $ T_p$; constraints: $ \mathcal{D}$, $\alpha_a$, $\alpha_e$, $ H_{\text{safe}}$, $ H_{\max}$\\
\textbf{GA (Global joint optimization):} \\
1: \textbf{for} $i=1$ to $N_g$ \textbf{do}\\
2:\hspace{1em} Randomly initialize $\mathbf{g}^{(0)}_i$ s.t. constraints; $F^{(0)}_i \leftarrow$ \eqref{eq:fitness}\\
3: \textbf{end for}\\
4: \textbf{for} iteration $t=0$ to $T_g\!-\!1$ \textbf{do}\\
5:\hspace{1em} \textbf{Selection:} sample $\mathcal{T}_l$;\  $\hat{\mathbf{g}}_a,\hat{\mathbf{g}}_b=\arg\max_{\mathbf{g}\in\mathcal{T}_l} F^{(t)}(\mathbf{g})\footnotesize$\\
6:\hspace{1em} \textbf{Recombination:} form new candidates $\mathbf{c}_1,\mathbf{c}_2$ via \eqref{eq:crossover}\\
7:\hspace{1em} \textbf{Mutation:} with prob. $p_{\text{mut}}$, apply \eqref{eq:mutation} $ \forall j$\\
8:\hspace{1em} \textbf{Repair:} $\mathbf{c}_q \leftarrow {R} (\mathbf{c}_q)$; evaluate $F(\mathbf{c}_q)$ via \eqref{eq:fitness}, $q=1,2$\\
9:\hspace{1em} \textbf{Elitism:} $\{\mathbf{g}^{(t+1)}_i\}\!\leftarrow\!\mathrm{top}_{\mu_{\text{elite}}}\big(\{\mathbf{g}^{(t)}_i\}\!\cup\!\{\mathbf{c}_1,\mathbf{c}_2\}\big)$\\
10:\hspace{0.3em}\textbf{end for}\\
11: Assign $\mathbf{g}^\ast \leftarrow \arg\max_{\mathbf{g}\in \mathbf{g}^{(T_g)}_i} F(\mathbf{g})$\\
\textbf{PSO (Per-UAV local refinement):}\\
12: \textbf{for} $m=1$ to $M$ \textbf{do}\\
13:\hspace{1em} Init $\{\mathbf{s}^{(0)}_{i,m}\}_{1}^{N_p}$; $\mathbf{s}^{(0)}_{1,m}\!\leftarrow[\mathbf{g}^{\ast}]_{m}$; others $\sim\mathcal{U}(\text{constraints})$\\
14:\hspace{1em} Init $\{\mathbf{v}^{(0)}_{i,m}\}$;\ $\mathbf{p^{best}_{i,m}}$ $\!\leftarrow\mathbf{s}^{(0)}_{i,m}$\\
15:\hspace{1em} $\mathbf{g^{best}_m}\!\leftarrow\arg\max_i F(\mathbf{s}^{(0)}_{i,m}\mid\text{others fixed})$\\
16:\hspace{1em} \textbf{for} $t=0$ to $T_p\!-\!1$ \textbf{do}\\
17:\hspace{2em} $w^{(t)} \leftarrow w_{\max} - \frac{w_{\max}-w_{\min}}{T_p}\,t$\\
18:\hspace{2em} \textbf{for} $i=1$ to $N_p$ \textbf{do}\\
19:\hspace{3em} Draw $r_1,r_2\sim\mathcal{U}(0,1)$; update $\mathbf{v}^{(t+1)}_{i,m}$ via \eqref{eq:pso-vel}\\
20:\hspace{3em} $\mathbf{s}^{(t+1)}_{i,m}\!\leftarrow{R}\big(\mathbf{s}^{(t)}_{i,m}+\mathbf{v}^{(t+1)}_{i,m}\big)$ using \eqref{eq:pso-pos}\\
21:\hspace{3em} Evaluate $F^{(t+1)}_{i,m}\!=\!F(\mathbf{s}^{(t+1)}_{i,m}\mid \text{others fixed})$\\
22:\hspace{3em} Update $\mathbf{p^{best}_{i,m}}$, $\mathbf{g^{best}_m}$ if improved\\
22:\hspace{2em} \textbf{end for}\\
23:\hspace{1em} \textbf{end for}\\
24:\hspace{1em} Set refined UAV-$m$ state to gbest$_m$\\
25:\textbf{end for}
\vspace{0.2em}
\textbf{Output:} Final deployment $\widehat{\mathbf{g}}=\{\text{gbest}_m\}_{m=1}^{M}$.
\label{algo:ga_pso}
\end{algorithm}
\vspace{-0.8cm}

\begin{equation}
\footnotesize
F^{(t)}_{i} = \lambda_{S}\,P_{\mathrm{sum}}\!\big(\mathbf{g}^{(t)}_{i}\big) - \lambda_{E}\,E_{\mathrm{total}}\!\big(\mathbf{g}^{(t)}_{i}\big) - \lambda_{\mathrm{pen}}\sum_{j}\mathrm{Violation}^{(t)}_{i,j},
\vspace{-0.2cm}
\label{eq:fitness}
\tag{17}
\end{equation}
where $\lambda_{\mathrm{pen}}$ is the penalty coefficient, and $\mathrm{Violation}^{(t)}_{i,j}$ quantifies the $j$-th constraint violation for solution $i$ at generation $t$. This penalty term steers the search toward feasible regions, ensuring that constraint requirements are satisfied.

Candidate selection employs tournament sampling (Line-5), where a subset $\mathcal{T}_l$ of $l$ solutions is drawn without replacement, and the best solution is chosen as $\hat{\mathbf{g}} = \arg\max_{u \in \mathcal{T}_l} F^{(t)}_{u}$. This balances exploitation of strong solutions with diversity introduced through random sampling. 

Recombination is applied by uniformly mixing two selected solutions (Line-6), $\mathbf{g}^{(t)}_{a}$ and $\mathbf{g}^{(t)}_{b}$, to generate new candidates $\mathbf{c}_{1}$ and $\mathbf{c}_{2}$. For each component index $j$, the update rule is
\begin{equation}
\vspace{-0.1cm}
\footnotesize
\mathbf{c}_{1}[j] =
\begin{cases}
\mathbf{g}^{(t)}_{a}[j], & u_j < 0.5,\\
\mathbf{g}^{(t)}_{b}[j], & \text{otherwise},
\end{cases}
\quad
\mathbf{c}_{2}[j] =
\begin{cases}
\mathbf{g}^{(t)}_{b}[j], & u_j < 0.5,\\
\mathbf{g}^{(t)}_{a}[j], & \text{otherwise},
\end{cases}
\label{eq:crossover}
\tag{18}
\end{equation}
where $u_j \sim \mathcal{U}(0,1)$ is a uniform random variable. To maintain search diversity, mutation perturbs solution components with probability $p_mut$, given by
\begin{equation}
\mathbf{g}^{(t)}_{i}[j] \leftarrow \mathbf{g}^{(t)}_{i}[j] + \delta_j, 
\quad \delta_j \sim \mathcal{U}(-\Delta_j,\Delta_j),
\label{eq:mutation}
\tag{19}
\end{equation}
where $\Delta_j$ specifies the allowable perturbation step for component $j$ (Line-7). After recombination and mutation, the repair operator ${R}(\cdot)$ restores feasibility (Line-8), while elitism preserves the top $\mu_{\text{elite}}$ solutions across generations (Line-9). The best deployment obtained after $T_g$ iterations, denoted $\mathbf{g}^*$, is passed to the PSO refinement stage (Line-11).

\textbf{PSO Stage:}
Here, each UAV $m$ is refined independently using $N_p$ particles, with the other UAV states fixed at $\mathbf{g}^*$. The state of particle $i$ at iteration $t$ is $\mathbf{s}^{(t)}_{i,m} = [x^{(t)}_{i,m},y^{(t)}_{i,m},z^{(t)}_{i,m},\eta^{(t)}_{i,m},\zeta^{(t)}_{i,m}]$, and its initial velocity $\mathbf{v}^{(t)}_{i,m}\in\mathbb{R}^5$ is assigned randomly to promote variability, while the first particle is initialized at $\mathbf{g}^*$ (Line-13). The particle evaluation mirrors the GA fitness in \eqref{eq:fitness}, which is penalty-based, but with only UAV $m$ allowed to vary, i.e., $F^{(t)}_{i,m}=F^{(t)}_{i}(\mathbf{s}^{(t)}_{i,m}\mid\text{others fixed})$ (Line-15).

Each UAV $m$ updates its state iteratively by balancing personal experience $(p^{best}_{i,m})$ and collective experience $(g^{best}_{m})$. The inertia weight $w$ is scheduled linearly from $w_{\max}$ to $w_{\min}$ (Line-17), enabling broad exploration in the early iterations and stable convergence as $T_{p}$ is approached. The velocity of particle $i$ at iteration $t$ is updated as
\begin{equation}
\small
\mathbf{v}^{(t+1)}_{i,m} = w\,\mathbf{v}^{(t)}_{i,m} + c_{1}r_{1}\big(\mathbf{p^{best}_{i,m}}-\mathbf{s}^{(t)}_{i,m}\big) + c_{2}r_{2}\big(\mathbf{g^{best}_{m}}-\mathbf{s}^{(t)}_{i,m}\big),
\label{eq:pso-vel}
\tag{20}
\end{equation}
where $c_{1}$ and $c_{2}$ are acceleration coefficients, and $r_{1},r_{2}\sim \mathcal{U}(0,1)$ are independent uniform random variables (Line-19). The first term preserves momentum, the second term attracts the particle toward its personal best, and the third term drives it toward the global best. The updated state is then computed as
\begin{equation}
\mathbf{s}^{(t+1)}_{i,m} = {R}\!\left(\mathbf{s}^{(t)}_{i,m} + \mathbf{v}^{(t+1)}_{i,m}\right),
\label{eq:pso-pos}
\tag{21}
\end{equation}
where $R(\cdot)$ enforces feasibility by projecting solutions as per the constraints (16a) to (16e) (Line-20). The best states $\mathbf{p^{best}_{i,m}}$ and $\mathbf{g^{best}_m}$ are updated when improved (Line-22), and after $T_p$ iterations, $\mathbf{g^{best}_m}$ is adopted as the refined placement for the $m^{th}$ UAV (Line-25).

\vspace{-0.1cm}
\section{Numerical Results}
\vspace{-0.15cm}

We evaluate the proposed hierarchical GA+PSO optimization framework through Monte Carlo simulations in Python~(v3.12.3) using a real-world 3D terrain model. A $5~\mathrm{km} \times 5~\mathrm{km}$ area with bottom-left corner at $(76.6479^{\circ}\mathrm{E},\ 34.0449^{\circ}\mathrm{N})$ was extracted from the Shuttle Radar Topography Mission (STRM) GeoTIFF dataset (1$^{\prime\prime}$, i.e., one arcsecond $\approx$ 30 m resolution)~\cite{USGS_SRTM} and approximated by 50 Gaussian components through curve fitting, achieving a training RMSE of $13.61$ $m$. This compact representation preserves the terrain profile while enabling efficient BVH-based line-of-sight verification. The targets are located at $\mathbf{p^t}_{1} = (3000, 3600, 5900)\,\text{m}$, $\mathbf{p^t}_{2} = (2000, 2300, 5800)\,\text{m}$, and $\mathbf{p^t}_{3} = (3800, 1100, 5700)\,\text{m}$ with transmission frequency sets $F_1 = \{105,110,\dots,245\}$, $F_2 = \{205,210,\dots,345\}$, and $F_3 = \{305,310,\dots,395\}$ MHz, each at 20 dBm. UAVs are deployed in a $1750 \times 5000\,\text{m}^2$ region, with altitudes initialized within bounds by (15e). The simulation parameters are: $K=1000$, $P_{\mathrm{fa}}=0.001$, beam width $(\alpha_a,\alpha_e)=(15^{\circ},15^{\circ})$, $N_t=7$, $\beta_0=-20$ dB, $\sigma_n^2=-80$ dBm, $S_{\min}=500$ m, $R_{\min}=200$ m, $H_{\mathrm{safe}}=50$ m, $H_{\max}=6000$ m, $P_0=275.204$ W, $H_s=44330$ m, and $t_d=900$ s. For BVH-adaptive LoS determination, $k_0=2$ and $\varepsilon$ is $10^{-5}$. The scalarization weights are set to $\lambda_S=2$ and $\lambda_E=5\times 10^{-3}$. The GA employs $N_g=50$, $p_{\mathrm{mut}}=0.1$, $\mu_{\mathrm{elite}}=5$, $T_g=100$, while PSO refines each UAV with $N_p=30$, $T_p=50$, $(w_{\max}, w_{\min})=(0.7,0.4)$, coefficients $(c_1,c_2)=(1.5,2.0)$, and penalty $\lambda_{\mathrm{pen}}=10^6$.

The proposed framework is evaluated against two baseline schemes. In the \textit{non-optimized scheme}, UAVs are deployed close to the targets and raise their altitude until LoS is achieved, after which antenna orientations are refined using a block coordinate descent algorithm. In the \textit{PSO-based scheme} \cite{Moliya2025}, a per-UAV PSO algorithm is applied, where each UAV optimizes its position and antenna orientation independently using a penalty-based fitness function to handle constraints.

Fig.~\ref{fig:UAV_Deployment_GA+PSO} shows the optimized UAV deployment strategy obtained using the proposed hierarchical GA+PSO framework across different UAV counts. In Fig.~\ref{fig:UAV_Deployment_GA+PSO}a, the UAV assigned to the sensing band [100, 200] MHz is placed close to Target~1 along the edge of the deployable region while lowering its altitude to reduce hover energy consumption, yet still maintaining line-of-sight connectivity. Conversely, in Fig.~\ref{fig:UAV_Deployment_GA+PSO}c, the UAV operating in the [160, 220] MHz band for Targets~1 and~2 increases its altitude to shorten the effective path length, thereby enhancing the received SINR and maximizing the cooperative sum detection probability. These adaptations show the framework’s use of altitude diversity to balance energy efficiency and detection reliability under constraints.

\begin{figure}[t]
    \centering
    \includegraphics[width=0.42\textwidth]{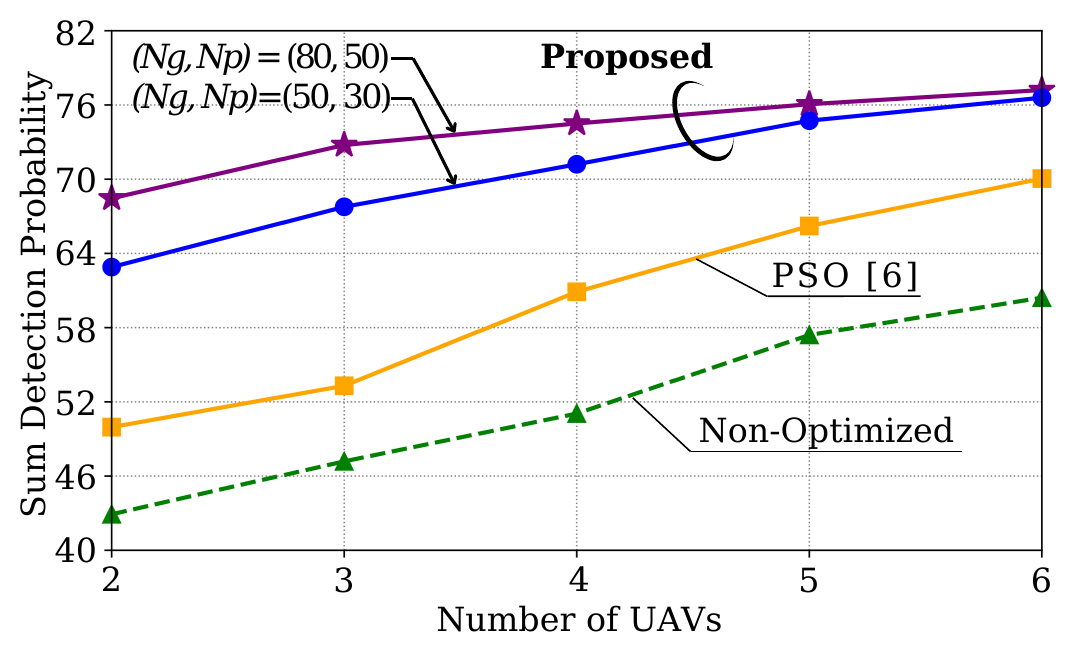}
    \vspace{-0.45cm}
    \caption{Comparison of sum detection probabilities with varying numbers of UAVs deployed across different schemes.}
    \vspace{-0.35cm}
    \label{fig:sum_detection_probability_plot}
\end{figure}

\begin{figure}[t]
    \centering
    \includegraphics[width=0.42\textwidth]{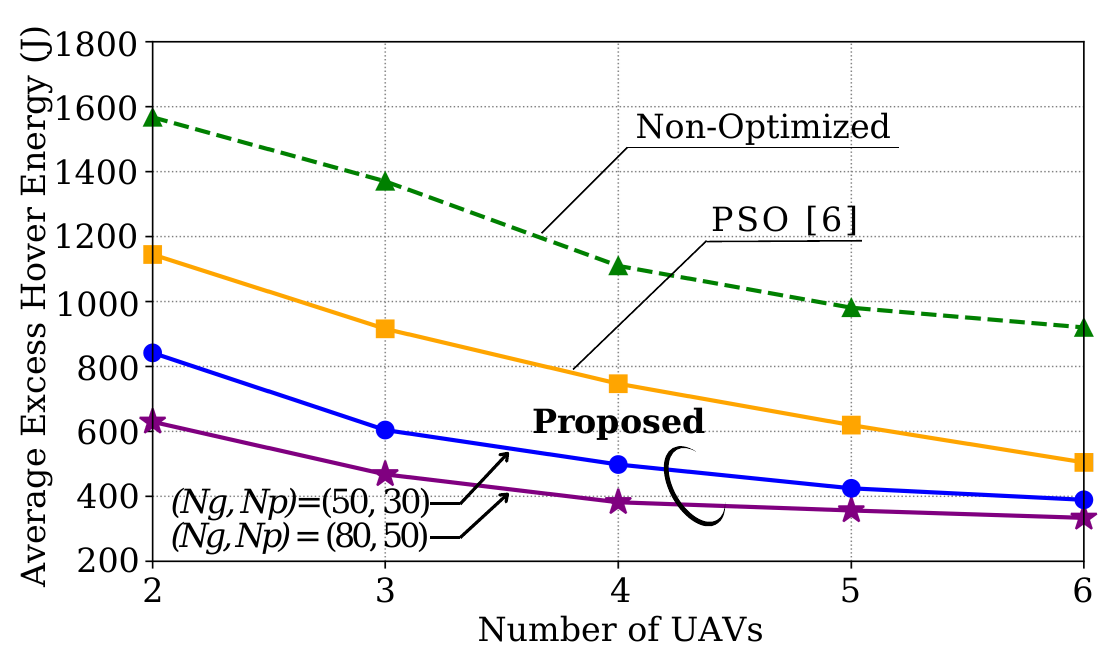}
    \vspace{-0.45cm}
    \caption{Comparison of average excess hover energy per UAV with numbers of UAVs deployed across different schemes.}
    \label{fig:avg_Excess_hover_energy_plot}
    \vspace{-0.55cm}
\end{figure}

To ensure statistical reliability, each experiment was repeated over 100 Monte Carlo runs with 95\% confidence intervals computed for all detection metrics. As shown in Fig.~\ref{fig:sum_detection_probability_plot}, the confidence intervals for UAV counts of 2 to 6 are ±2.14, ±1.46, ±1.38, ±1.32, and ±1.22, respectively. The narrowing intervals indicate improved reliability with more UAVs, though the plateau suggests diminishing returns beyond a certain count. For performance comparison, targets were randomly positioned, and each scheme was applied to optimize UAV deployment and antenna orientation. Results in Fig.~\ref{fig:sum_detection_probability_plot} Show that the proposed framework outperforms both baseline schemes, achieving a higher detection probability of up to 37. 0\% and 36. 5\% for 2 and 3 UAVs over PSO~\cite{Moliya2025}, and 59. 5\% and 54. 2\% over nonoptimized deployment, due to effective SINR optimization, balanced exploration–exploitation, and terrain-aware LoS handling over uneven terrain.

Fig.~\ref{fig:avg_Excess_hover_energy_plot} compares the average excess hover energy per UAV across different UAV counts. The proposed framework achieves substantially lower energy consumption, with reductions of 45.01\% for 2 UAVs and 48.94\% for 3 UAVs relative to PSO~\cite{Moliya2025}, and 59.84\% and 65.87\% relative to the non-optimized scheme. These gains result from the hierarchical strategy that combines global exploration via GA with per-UAV exploitation through PSO, enabling efficient navigation of the non-convex search space while maintaining terrain-aware LoS links and positioning UAVs at energy-efficient altitudes without compromising sensing reliability. Larger population sizes $(N_g, N_p)$ can further improve performance and robustness but increase computational cost, requiring a trade-off between accuracy and efficiency.

\vspace{-0.15cm}

\section{Conclusion}

\vspace{-0.05cm}
We presented in this paper a novel LoS-aware 3D deployment framework for multi-UAV CSS over uneven terrain, jointly optimizing UAV positions and antenna orientations as a bi-objective formulation that maximizes detection probability while minimizing hover energy. To efficiently incorporate binary terrain-aware LoS constraints, we propose a BVH-based adaptive indicator, and to address the complex non-convex problem, we introduced a hierarchical heuristic framework that combines global exploration through GA with per-UAV refinement via PSO. Monte Carlo simulations over a real-world SRTM-based uneven terrain model validated the effectiveness of the proposed approach, showing detection probability improvements of up to 37.0\% and 36.5\% (for 2 and 3 UAVs) over PSO~\cite{Moliya2025}, while simultaneously achieving energy reductions of 45.0\% and 48.9\% relative to PSO~\cite{Moliya2025}. These results highlight the importance of terrain-aware deployment for achieving reliable and energy-efficient UAV networks in realistic environments.

\vspace{-0.15cm}

\section*{Acknowledgment}
\vspace{-0.1cm}
The authors acknowledge the support provided by the Machine Intelligence, Computing, and xG Networks (MICxN) Lab and the Stepwell High Performance Computing Facility at the School of Engineering and Applied Science, Ahmedabad University.

\end{document}